# Sex differences in predicting fluid intelligence of adolescent brain from T1-weighted MRIs

Sara Ranjbar [0000-0002-4344-1282], Kyle W. Singleton [0000-0002-9969-8474], Lee Curtin [0000-0002-4083-3803], Susan Christine Massey [0000-0002-4680-7796], Andrea Hawkins-Daarud [0000-0002-6391-4155], Pamela R. Jackson [0000-0002-8424-4007], and Kristin R. Swanson [0000-0002-2464-6119]

Mathematical NeuroOncology Lab, Precision Neurotherapeutics Innovation Program, Department of Neurological Surgery, Mayo Clinic, Phoenix, AZ, USA
`ranjbar.sara@mayo.edu`

**Abstract.** Fluid intelligence (Gf) has been defined as the ability to reason and solve previously unseen problems. Links to Gf have been found in magnetic resonance imaging (MRI) sequences such as functional MRI and diffusion tensor imaging. As part of the Adolescent Brain Cognitive Development Neurocognitive Prediction Challenge 2019, we sought to predict Gf in children aged 9-10 from T1-weighted (T1W) MRIs. The data included atlas–aligned volumetric T1W images, atlas–defined segmented regions, age, and sex for 3739 subjects used for training and internal validation and 415 subjects used for external validation. We trained sex-specific convolutional neural net (CNN) and random forest models to predict Gf. For the convolutional model, skull-stripped volumetric T1W images aligned to the SRI24 brain atlas were used for training. Volumes of segmented atlas regions along with each subject's age were used to train the random forest regressor models. Performance was measured using the mean squared error (MSE) of the predictions. Random forest models achieved lower MSEs than CNNs. Further, the external validation data had a better MSE for females than males (60.68 vs. 80.74), with a combined MSE of 70.83. Our results suggest that predictive models of Gf from volumetric T1W MRI features alone may perform better when trained separately on male and female data. However, the performance of our models indicates that more information is necessary beyond the available data to make accurate predictions of Gf.

## 1 Introduction

Fluid intelligence (Gf) is the ability to reason and solve previously unseen problems [1]. It is highly associated with general intelligence, more so than the other intelligence subtypes [2] and has been linked to academic performance [3]. It is generally theorized that Gf is highly dependent upon biological processes and as such is more independent from previous learning than other types of intelligence, such as crystalized intelligence [4]. Gf is known to increase throughout childhood, peak around early adulthood and then decrease throughout adult life [4–7]. Fluid intelligence is typically measured using a battery of tests evaluating aspects of memory and pattern recognition [4].

Magnetic resonance imaging (MRI) is a powerful tool for non-invasively visualization of the body's tissues. Structural MRI, including T1-weighted (T1W) and T2-weighted scans, exhibit excellent contrast for non-invasively discriminating many brain structures. Advanced and quantitative MRI can provide more than just structural information, including brain activity as detected by functional MRI (fMRI) and white matter tractography from diffusion tensor imaging (DTI). Previous studies have shown that fluid intelligence can be predicted by imaging, in particular fMRI [2, 8] and white matter tract diffusion tensors from DTI [6]. In addition, levels of N-acetylaspartate and brain volume from MRI spectroscopy were shown to be associated with aspects of Gf, but not Gf itself [9]. To our knowledge, fluid intelligence has not been previously linked to T1W imaging.

Discovering connections between cognitive traits and non-invasive biomedical imaging could prove to be important for further understanding the neural underpinnings of cognitive development. Recently deep learning models trained on brain MRI data have shown promising results in the diagnosis of Alzheimer's disease [10], prediction of age [11, 12] and classification of overall survival in brain tumor patients [8]. In this work we compared the performance of deep learning techniques with that of classic machine learning techniques trained on hand-crafted features in the prediction of Gf from MRI-based features. We also opted to train separate models for male and female cohorts. The potential impact of biological sex differences is a recommended consideration for neuroscience research [13]. Overall, males have been reported to have larger brains, and in young subjects who are still developing, sex differences have been reported in the growth and maturation of various brain structures [14, 15]. Consequently, sex differences in the developmental state of brain structures may affect MRI features, including the size and texture of various brain regions. It is not yet clear if these factors are related to Gf. By conducting our analysis in a sex–specific manner, we sought to reduce any erroneous association between the development of brain structures and Gf.

## 2    Material

Data used in this work was acquired by the Adolescent Brain Cognitive Development (ABCD) study (abcdstudy.org) and access was provided to participants as part of the ABCD Challenge (sibis.sri.com/abcd-np-challenge). The ABCD study is the largest long-term study of brain development and child health in the United States. The overall challenge cohort included data from 4154 children ages 9-10. Extensive information about ABCD data can be found in the Data supplement of [16]. Data was split into 3739 subjects in the training set (47.4% of which were females) and 415 (49.3% females) in the external validation set. The provided skull-stripped T1W MRIs underwent pre-processing for conversion to the NifTI format, noise removal, and field inhomogeneity correction as described in [17]. Further, all volumetric MRIs were standardized to a 240x240x240 dimensionality with voxel resolution of 1mm in x, y, and z directions.

The images were also affinely aligned to the SRI 24 atlas [18] and segmented into cortical and subcortical structures (e.g. parcellated gray matter, white matter, and cerebral fluid) according to the atlas. Volumetric scores of 122 segmented brain regions, along with participants' sex and age at interview in months were provided to participants in the format of a csv file. Accompanying fluid intelligence scores were assessed using a variety of tests as detailed in [4]. Scores were normalized using a linear regression model trained on factors such as brain volume, data collection site, age at baseline, sex at birth, race/ethnicity, highest parental education, parental income, and parental marital status. Demographic confounding factors such as sex and age were removed from the scores by the ABCD study. Fig. 1 shows the distribution of fluid intelligence across the training and validation sets for males and females.

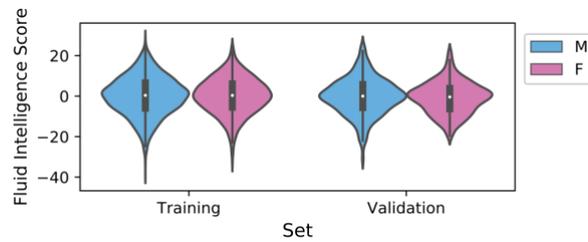

**Fig. 1.** Distribution of fluid intelligence scores across training and validation sets for males and females.

## 3   Methods

We conducted two separate experiments to predict fluid intelligence scores: in the first, we used skull-stripped T1W images as the input to train 3D convolutional neural nets to predict Gf scores. In the second, we trained random forest regression models using the volumetric features from image-atlas alignment along with the age and sex of subjects. For both experiments, we trained models on sex–specific cohorts and predicted the Gf of sex-equivalent subjects in the validation set. Male and female predictions were used to compute sex-specific MSE scores, and then a total population MSE was calculated using the merged set of sex-specific predictions. For a comparative baseline performance, we created a predictor with zero rule algorithm using the mean Gf in the training cohort and measured the model accuracy in predicting Gf for the validation set.

### 3.1   Convolutional Neural Networks

We used residual convolutional neural networks in the first experiment. The residual architecture (ResNet) has previously been described in [11]. Nets were trained on a Nvidia TITAN V GPU using Keras with TensorFlow backend running through Nvidia docker, Python 3, and TensorFlow 1.12.0. To adjust for the computational capabilities of our GPU, we resized the images to 128x128x128 voxels and used a stride of 2 in the initial convolutional layer. Several areas of the brain, such as the caudate and putamen, have previously been reported to have links to fluid intelligence [19]. As a result, we repeated training with the ResNet using only slices that included the caudate and putamen (i.e., slices 55-75) resulting in image dimensionality of 128x128x20. Training was performed using the RMSprop optimization algorithm for learning the weights, a batch size of 32, an MSE loss function, and 100 epochs. Learning rate was initially set to 0.001 and was reduced by a factor of 0.1 at validation loss plateau.

As mentioned before, separate models were generated for male and female cohorts. A total of 1966 MRIs were available for the male model and 1773 MRIs for the female model. Prior to training, the sex-specific training sets were further split into training and internal validation sets using an 80:20 ratio. We did not perform data augmentation in this work.

### 3.2 Random Forests

In the second experiment we trained random forest regression models using volumetric and demographic features described in Section 2. We used the scikit-learn [20] package and Python 3.6.8 for this experiment. Hyperparameter selection was performed using grid search for maximum depth (range of 2-6) and the number of estimators in the random forest (100-500, 750, and 1000 trees). The remaining hyperparameters were left at default values in scikit-learn. The same set of 1966 male and 1773 female subjects were used in these sex-specific models.

## 4 Results

Males and females shared a median age of 120 months in training and validation sets. The ratio of males to females was 1.11:1 in the training set and 1.02:1 in the validation set for a total of 1.10:1 for the entire cohort.

### 4.1 CNN Analysis Results

Overall performance of the CNN attempts was poor. Sex-specific ResNet using the complete T1W MRIs had better total performance than the caudate-putamen T1W slices alone. However, MSE was much lower in female cases in the caudate-putamen case, indicating possible influence of sex on those regions. None of the CNN models evaluated in our analysis was able to obtain a lower MSE than the baseline predictor.

### 4.2 Random Forest Analysis Results

Table 1 shows the top-10 most important features to the male and female random forest models. Pons white matter volume was most important for both sexes, which may highlight the important role of sensory processing carried out by the pons [21]. Yet, there are notable differences in the selected features between the two models. In particular, hippocampus volumes (both left and right) were important among males but not females. Since the hippocampus is generally considered to play a pivotal role in the consolidation of short-term to long-term memory through detection of new stimuli [22], and is also active in navigation and spatial memory [23, 24], it is striking that it only appeared to be of considerable importance in one sex. Overall, random forest models utilizing demographic and brain region volume data had the best performance in our analysis with a smaller MSE than the baseline predictor.

**Table 1.** Top 10 important features along with their importance scores for male and female random forest models. For further information on what these features represent the avid reader is referred to the supplement of [16].

| Male | | Female | |
|---|---|---|---|
| **Feature** | **Importance** | **Feature** | **Importance** |
| ponsWM | 0.042 | ponsWM | 0.0523 |
| hippocampus_L_GM | 0.041 | cerebelum8_R_VOL | 0.0391 |
| cerebelum6_R_VOL | 0.0327 | temporalinf_L_GM | 0.0282 |
| cerebelum45_R_VOL | 0.0316 | WM400WM400_R_WM | 0.0269 |
| hippocampus_R_GM | 0.0277 | fusiform_R_GM | 0.0209 |
| parietalinf_R_GM | 0.0254 | calcarine_R_GM | 0.0207 |
| frontalmid_L_GM | 0.0189 | temporalsup_L_GM | 0.02 |
| parahippocampal_L_GM | 0.0181 | precuneus_L_GM | 0.019 |
| amygdala_L_GM | 0.0181 | cuneus_R_GM | 0.017 |
| cerebelumcrus2_L_VOL | 0.0175 | cerebelum45_R_VOL | 0.0162 |

### 4.3 Comparison of CNN and Random Forest

Table 2 compares the performance of CNN and random forest models in prediction of Gf. ResNet model MSE values were significantly higher than the baseline predictor in a 2-sided t-test (ResNet Full Brain: p-value=0.0006; ResNet Caudate-Putamen: p-value=7.9e-5). The difference in the MSE for random forest models from the baseline predictor was lower by 1.01, but this difference was not significantly different from the baseline predictor (p-value = 0.17). The female random forest model presented a lower MSE than the male model, similar to the trends observed for ResNet models. Fig. 2 compares the distribution of predicted Gf scores all models. As evident, all models failed to predict the full range of Gf scores in the validation set and yielded predictions close to the average Gf.

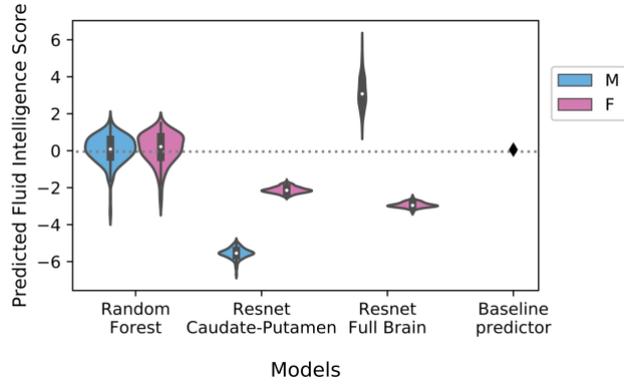

**Fig. 2.** Distributions of predicted fluid intelligence scores in our experiments. Dashed line shows the average fluid intelligence score in the validation set.

**Table 2.** Summary of the performance of models on internal/external validation sets

| | | External Validation MSE score | |
|---|---|---|---|
| Classification method | Sex (Training) | Sex Specific | Combined Population |
| Resnet (Full Brain) | M | 93.57 | 79.23 |
| Resnet (Full Brain) | F | 77.12 | |
| Resnet (Caudate-Putamen) | M | 111.83 | 87.05 |
| Resnet (Caudate-Putamen) | F | 61.67 | |
| Random Forests | M | 80.74 | 70.83 |
| Random Forests | F | 60.68 | |
| Baseline predictor | - | - | 71.84 |

## 5  Discussion

Fluid intelligence has been an established metric in psychology and education research since the early 70's [25]. Standard Gf assessments primarily rely on non-verbal multiple choice questionnaires [26]. While there have been a few studies investigating whether Gf scores are predictable from functional MRI, to date, there have been no studies investigating possible connections from structural T1W MRI. As part of the ABCD challenge, we present a new line of investigation into whether anatomical attributes of an individual's brain as visualized on a T1W MRI are predictive of Gf.

In this work, we investigated this hypothesis by assessing the performance of different machine learning models in predicting Gf. Convolutional neural network models were trained on slices presenting selected brain structures (caudate and putamen) as well as on the entire volume of T1W MRI images. Random forest regressor was trained using the volumes of different brain regions. We focused on sex-specific versions of these models due to previously observed sex-differences in structural MRI in the growth and maturation of various brain structures in children and adolescent subjects [14, 15]. We chose to consider both the CNNs and the random forest as they have different strengths at evaluating the available data. Random forest is a powerful machine learning method capable of incorporating both categorical and numerical hand-crafted features in the decision-making process. CNNs on the other hand, do not require hand-crafted features and are able to extract imaging patterns at both small and large scales. Since none of the models presented a compelling accuracy separately, we did not merge the two into a combined model.

Given the large size of the ABCD dataset, it was somewhat surprising that CNN was outperformed by the random forest model. However, a closer look at the predictions indicated that they were all tightly clustered around one or another Gf score (Fig. 2) demonstrating that all models failed to predict the full range of intelligence scores. The random forest model marginally described the range of Gf scores and therefore achieved a lower MSE than CNNs. This strongly suggests that the relationship between Gf and imaging features seen on T1W MRI is not predictive. It was also somewhat unexpected that the female model outperformed the male model. The large difference in MSE between the female and male models is likely attributable to the reduced number of outlier Gf scores in the female validation subcohort (Fig. 1). However, the fact that volumes of different brain regions contributed to the two models (Table 1) may indicate sex differences in the underlying mechanisms of fluid intelligence in developing brain. Further investigation is needed to determine whether there are in fact different imaging predictors of Gf among developing young male vs. female brain.

Ultimately, T1W MRIs and volumetric features did not provide a compelling accuracy in prediction of fluid intelligence scores in our analyses. While it is possible that our approaches were not sophisticated enough to detect the predictive relationship, the similarity of all approaches to a baseline zero rule algorithm using the training set mean suggests that the characteristics of physical brain structures as identifiable on T1W MRI do not explain Gf, despite being individualized. Other MRI sequences that have been reported to link with Gf, such as fMRI [2, 8] and DTI [6], contain more functional information than T1W MRI, suggesting that additional information is required to capture a relationship between structural imaging and fluid intelligence. It should also be noted that the complexity of this proposed problem could be confounded by the Gf scoring system.

## 6     Conclusion

In this work we experimented with predicting fluid intelligence of adolescent brain using MRI and a set of machine learning techniques. Overall, the performance of the best model was not significantly superior to a baseline predictor using the average fluid intelligence score. We associate the uninspiring predictive performance of the models to the insufficiency of structural MRI in explaining the complexity behind fluid intelligence mechanism in developing brain.

**Acknowledgements**
The authors would like to thank the Challenge Organizers and ABCD Study Researchers for the opportunity to participate and utilize their data. We also thank Kevin Flores, Erica Rutter, and John Nardini for many helpful discussions. Further, we acknowledge the following funding sources: James S. McDonnell Foundation, U54CA210180, U54CA193489, 3U54CA193489-04S3, and U01CA220378.